\documentclass[runningheads,a4paper]{llncs}

\usepackage{amssymb}
\usepackage{graphicx}
\usepackage{algorithmic}
\usepackage{algorithm}
\usepackage{url}
\urldef{\mailsa}\path|{olfa.arfaoui, minyar.sassi}@enit.rnu.tn|
\newcommand{\keywords}[1]{\par\addvspace\baselineskip
\noindent\keywordname\enspace\ignorespaces#1}

\begin{document}

\mainmatter  % start of an individual contribution

\title{Mining Semi-structured Data}

\titlerunning{Mining Semi-structured Data}

\author{Olfa Arfaoui$^{1}$ \and Minyar Sassi Hidri$^{2}$}
\authorrunning{Mining Semi-structured Data}

\institute{$^{1,2}$Universit\'e de Tunis el Manar,\\
Ecole Nationale d'Ing\'enieurs de Tunis,\\
BP. 37, Le Belv\'ed\`ere 1002 Tunis, Tunisia\\
\mailsa}


\toctitle{Mining Semi-structured Data}
\tocauthor{Arfaoui, O., and Sassi-Hidri, M.}
\maketitle

\begin{abstract}
The need for discovering knowledge from XML documents according to both structure and content features has become challenging, due to the increase in application contexts for which handling both structure and content information in XML data is essential. So, the challenge is to find an hierarchical structure which ensure a combination of data levels and their representative structures. In this work, we will be based on the Formal Concept Analysis-based views to index and query both content and structure. We evaluate given structure in a querying process which allows the searching of user query answers.
\keywords{XML Mining, Formal Concept Analysis, Conceptual Scaling, XQuery, eXist}
\end{abstract}
\section{Introduction}
The widespread use of XML (eXtensible Markup Language) \cite{book1,url1} across the Web and in business as well as scientific databases has prompted the development of methodologies, techniques and systems for effectively and efficiently managing and analyzing XML data.

This has increasingly attracted the attention of different research communities, including databases (DB), information retrieval (IR), pattern recognition, and machine learning, from which several proposals have been offered to address problems in XML data management and knowledge discovery \cite{book4}.

XML Structure and Content Mining is one of these problems. It has its roots in problems which originally arose from several applications in semi structured data management, such as querying data sources and query processing.

Hence the recourse to the development of new indexing and querying systems whose aim is to provide a fast and a reliable XML data access because indexing technique influences the reliability of the querying process in terms of research time and treatment queries.

For this, several studies has been introduced \cite{proceeding1,proceeding2,proceeding3}. They belong to both the community of DB and the recent researches in XML language. The main purpose of these methods is to develop indexing approaches own to XML technology.

The main problem of these methods is how to find information in a document while taking into account structure and content. This presents a great challenge when we want mining large volumes of XML data. The structural dimension must be taken into consideration for the users's needs.

However, maintaining the hierarchical structure of elements of an XML document and their order is important to avoid recalculating each time this order and so avoid a sequential access to data in order to determine the relationship between elements.

The recourse to the Formal Concept Analysis (FCA) \cite{book2} to mining XML documents appears effective to find solutions to the indexing and querying problems while putting into account i) The extraction of the most representative words in the documents (key-words) and their structural information; ii) The structural aspect is assigned to the content and the following questions appear: How we can  index the document structure?, How we can connect this structure to the document content? and Depending on what dimension the indexing terms should be weighted?

To answer these questions, the contributions from this work should allow i) The reconstruction of the XML document decomposed in handling structures; ii) The processing of the path expressions on the XML structure; iii) The processing of precise predicates on the XML documents content and iv) The search data by key-words.

In this work, we  propose to summarize XML data into a conceptual scaling and  generate a generalized view seen as Concept Lattice, a FCA-based structure, to come up with the proposed mining method.

The rest of the paper is organized as follows: section 2 describes and evaluates our mining semi-structured data model. Section 3 concludes the paper and gives future work.
\section{Mining Semi-structured Data}
\subsection{Overview of our Mining XML Data Model}
Before querying XML data, we must proceed to index them. Recently, a new indexing approach have been proposed based on FCA and gives answers on several abstraction levels \cite{proceeding5}.

The main idea consists of using FCA based-theory on XML data in order to index them and subsequently facilitate the querying process of such data. The steps which compose our approach are:
\begin{itemize}
\item XML tree traversal: is to traverse the XML tree and extract the textual data in the form of a set $E$,
\item Conceptual classification: is to build the concept lattice associated to each parent nodes generated following an ascending traversal of the document,
\item Conceptual scaling: the lattice structures obtained are combined into a single structure called nested lattice base on conceptual scaling.
\end{itemize}

\begin{figure}[h!]
\centering
\includegraphics[height=6cm]{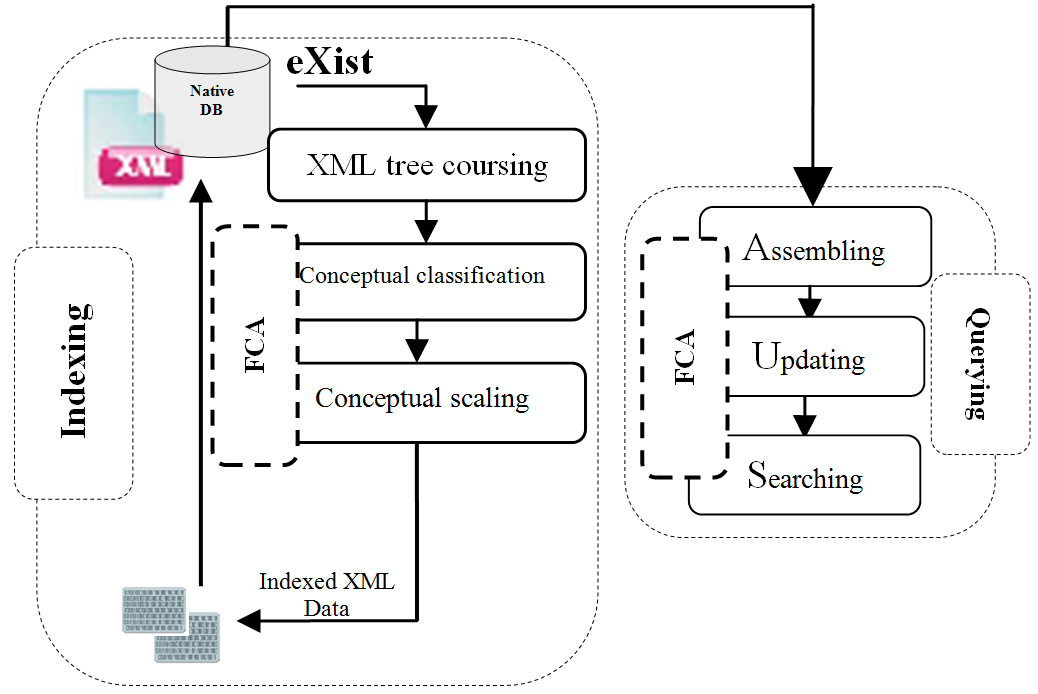}
\caption{Overview of the approach steps.}
\label{fig:1}
\end{figure}

After  indexing data, we proceed to the querying step which consists of three steps:
\begin{itemize}
\item Assembling: is to generalize the different concepts lattices into a generalized one.
\item Updating: is to transform a user query in a concept and insert it into the structure (concept lattice).
\item Coursing: is to course the concept lattice for generation of query answers.
\end{itemize}
Fig. \ref{fig:1} shows the overview of our approach.

\subsection{Indexing XML Data}
The overall process of indexing step is detailed in \cite{proceeding5}. From the nested concepts lattice (Conceptual Scaling) generated from the indexing step, we generalize the nested structure to generalized concept lattice.

\begin{figure}[h!]
\centering
\includegraphics[height=5cm]{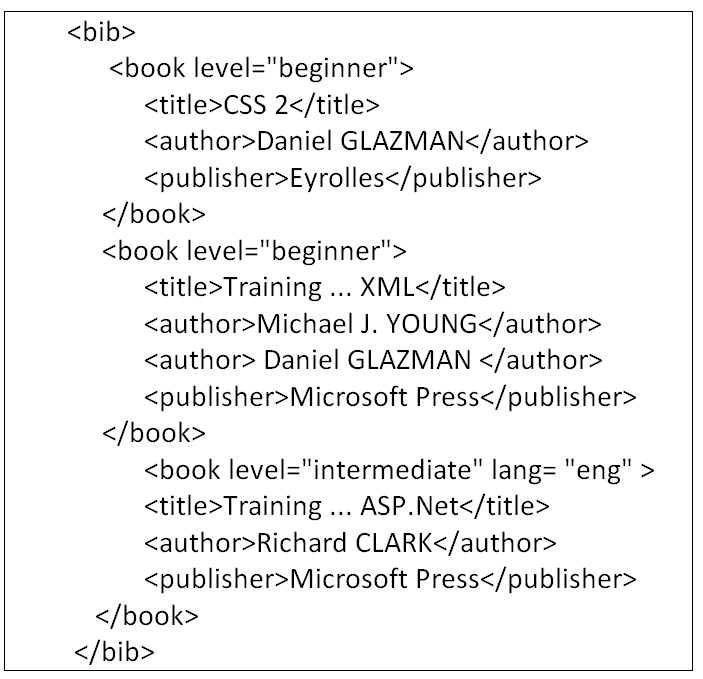}
\caption{Example of an XML tree.}
\label{fig:2}
\end{figure}

Fig. \ref{fig:2} shows an example of an XML document. The data that we may have in our example are extracted from leaves and nodes as follow: {\em Beginner}, {\em CSS 2}, {\em Daniel Glazman}, {\em Eyrolles}, {\em Training...XML}, {\em Michael J. YOUNG}, {\em Microsoft Press}, {\em Intermediate}, {\em Eng}, {\em Training ... ASP.Net}, {\em  Richard Clark}.

Let take $E=\{D_{1},D_{2},...,D_{11}\}$ the data set representing the leaf nodes of the XML tree.

The first step consists of extraction structure data of parent node. According to our example, the first book is a parent node and it has as structural data.

After extraction, Galois lattices associated to each parent node generated following an ascending traversal of the XML tree are constructed.

Table \ref{tab:1} shows an example of formal context of the XML data presented in \ref{fig:2}. The processed node is $<book>_{[0]}$.

For this node, the construction of the corresponding context witch consists of the set of words $E$ and the set of the child nodes $<level>$, $<title>$, $<author>$, $<publisher>$ can started. They represent
respectively the set of objects and attributes of this context.

\begin{table}[ht]
\caption{Binary context of the node $<book>_{[0]}$.} \label{tab:1}
 \begin{center}
 \small{
   \tabcolsep = 1\tabcolsep
   \begin{tabular}{cccccccccccc}
   \hline\hline
   $R$            & $D_{1}$  &$D_{2}$  & $D_{3}$ & $D_{4}$ & $D_{5}$ & $D_{6}$ & $D_{7}$ & $D_{8}$ & $D_{9}$ & $D_{10}$ & $D_{11}$ \\
   \hline
   $<level>$ & 1 & 0 & 0& 0& 0& 0& 0& 0& 0&0 &0 \\
   $<lang>$& 0 & 0 & 0& 0& 0& 0& 0& 0&0 & 0&0 \\
   $<title>$& 0 & 1 &0 &0 & 0& 0& 0&0 & 0& 0& 0\\
   $<author>_{[0]}$ &  0& 0 &1 & 0& 0& 0& 0&0 &0 & 0&0 \\
   $<author>_{[1]}$& 0 &0  &0 &0 &0 &0 &0 &0 & 0&0 &0 \\
   $<publisher>$ & 0 & 0 & 0& 1& 0&0 &0 &0 &0 &0 &0 \\
    \hline
   \end{tabular}
   }
 \end{center}
\end{table}

Similarly, $<book>_{[1]}$ and $<book>_{[2]}$ are defined. The nodes presented above have the same parent $<bib>$. So all the child nodes of these nodes become leaves.

Therefore in the context node $<bib>$, the lines represent the set of objects and the columns the set of attributes which are respectively the set of parent nodes $(<book>_{[0]}$, $<book>_{[1]}$, $<book>_{[2]})$ and all the leaf nodes $(<level>$, $<title>$, $<lang>$, $<author>$, $<author>$, $<Publisher>)$.
Table \ref{tab:2} illustrates this context.
\begin{table}[ht]
\caption{Binary context of the root node $<bib>$.} \label{tab:2}
 \begin{center}
   \small{
   \tabcolsep = 2\tabcolsep
   \begin{tabular}{cccc}
   \hline\hline
   $R$            & $<book_{[0]}>$  &$<book_{[1]}>$  & $<book_{[2]}>$  \\
   \hline
   $<level>$ & 1 & 1 & 1  \\
   $<lang>$& 0 & 0 & 1 \\
   $<title>$& 1 & 1 &1 \\
   $<author>_{[0]}$ &  1& 1 &1   \\
   $<author>_{[1]}$& 0 &1  &0  \\
   $<publisher>$ & 1 & 1 & 1  \\
    \hline
   \end{tabular}
   }
 \end{center}
\end{table}

The interest of a concept lattice is to organize information about groups of objects with common properties.

Taking the example of Table \ref {tab:2}, $(\{<book>_{[0]}>$, $<book>_{[1]}>$, $<book>_{[2]}>\}$, $\{<level>,$ $<title>\})$ and $(\{<book>_{[0]}>$, $<book>_{[1]}>$, $<book>_{[2]}>\}$, $\{<publisher>\})$ are both concepts.

The second concept means that objects $<book>_{[0]}$, $<book>_{[1]}$ and $<book>_{[2]}$ have in common the attribute $<publisher>$.

Several algorithms have been proposed for the construction of concept lattice. Their complexity is exponential and several techniques have been developed to reduce computation time \cite{proceeding6}.

\begin{figure}[h!]
\centering
\includegraphics[height=9cm]{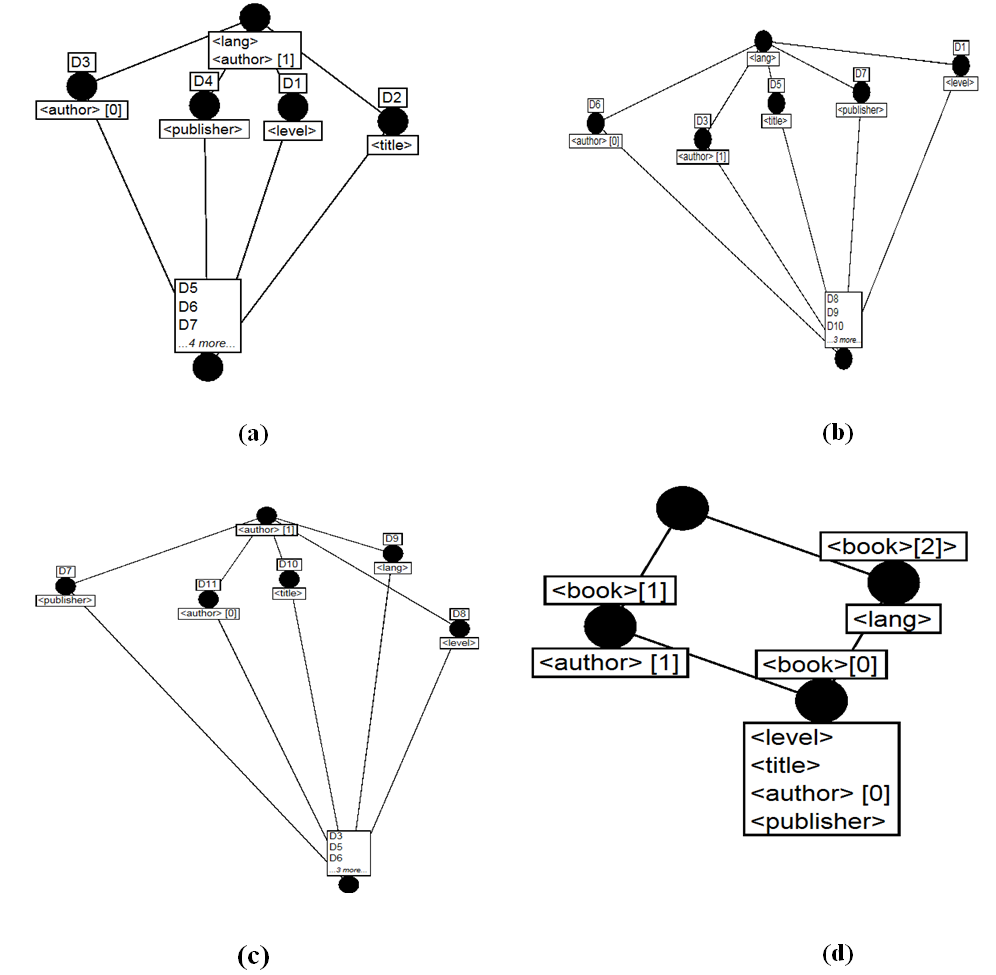}
\caption{Concepts lattice for $(a)<book>_{[0]}$, $(b)<book>_{[1]}$, $(c)<book>_{[2]}$ and $(d)<bib>$.}
\label{fig:3}
\end{figure}

Fig.\ref{fig:3} shows the concepts lattices for $<book>_{[0]}$, $<book>_{[1]}$, $<book>_{[2]}$ and $<bib>$ respectively.

If we want to index an XML document, it will be necessarily to focus on the content and structure. This is insufficient because the binary context is not satisfactory for indexing multi-valued attributes, hence the use of nested lattice based-structure (conceptual scaling), which aims to extend the interest of the simple Galois lattice.

This structure has been developed by Ganter and Wille \cite{book3}. The general process in the Conceptual Scaling begins with the representation of knowledge in a data table with arbitrary values and missing values probably. These data tables are formally described by multi-valued context $(G,M,W,I)$, where $G$ is a set of objects, $M$ is a set of {\em multi-valuated attribute}, $W$ is a set of values and $I$ is a ternary relation, $I\subseteq m\times g~W$, such that for all $g\in G$, $m\in M$ there is at most one value $w$ satisfying $(g,m,w)\in I$.

Therefore, a multi-valuated attribute $m$ is generally interpreted as a measurement function (partial) and we write $m(g)=w$ if and only if $(g,m,w)\in I$ such as presented in \cite{proceeding7}.
\subsection{Generalizing Concept Lattices}
From the nested concepts lattice (Conceptual Scaling) generated from the indexing step, we generalize the nested structure to generalized concept lattice. This lattice structure corresponds to the generalized view.

From this, we will be able to find answers to satisfy a given query. This step involves building the first concept associated with the user query. From this concept, we check the query feasibility. If it is, the concept is built and then inserted into the generalized view.

The answer to the query is then provided by the extraction of objects belonging to the upper bounds of extensions of concepts in the query concept.

For the construction of generalized view, we provide the following structural steps:
\begin{itemize}
\item Identification of global concepts:  characterization of full nodes of the nested lattice (Conceptual Scaling);
\item Calculation of the intention and extension of each concept;
\item Calculation of the hedging relationship of the lattice (immediate predecessors of a concept).
\end{itemize}
\subsection{Updating Generalized View}
To better explain the steps of evaluation, we consider an example of an XML data and we use XQuery language \cite{proceeding9}.

Consider the following query: {\em Returns a sequence of elements of type (publisher author) who are children of the first book element type}.

According to XQuery language, this query can be rewriting as follows:

{\em document(bib.xml)/bib/book}$_{[1]}$ {\em /(publisher, author);}

Once the generalized view is built, the search for answers can begin. For this, we define a query which is a concept. The concept extension is sought that all nodes of the XML document "bib.xml" and the set of objects sought by the query.

After that, the query is parsed and validated. These tokens are provided to {\em XQuery Analyzer}. It analyze tokens and create separate one path: {\em SearchPath}, {\em ConditionalPath} and {\em ReturnPath}. This is illustrating by the following query:

{\em For \$b in doc (bib.xml)/bib/book } $\leftarrow$ {\em SearchPath}

{\em Where \$b/author =Daniel GLAZMAN}$\leftarrow$ {\em ConditionalPath}

{\em Return \$b/book} $\leftarrow$ {\em ReturnPath};

where:

\begin{itemize}
\item {\em SearchPath()}: is a process which returns the elements in the path specified in the FOR clause.
\item {\em ConditionalPath()}: is a process which returns the subset of elements returned by {\em processForPath()} satisfying the condition in the {\em WHERE} clause of the XQuery query.
\item {\em ReturnPath()}: is a process which returns the elements (usually descendants of the  elements returned by {\em ConditionalPath()} process specified by the {\em RETURN} clause.
\end{itemize}

The set of attributes is determined by the following algorithm:\\
\begin{algorithm}[h!]
\small{
\caption{Construction of query concept}
\label{alg4}
\begin{algorithmic}
\REQUIRE XQuery query $Q$
\ENSURE Query concept $Q=(Q_{A},Q_{B})$
\STATE Read the query
\STATE Check for grammatical error
\IF {No-Error-Found}
\STATE Create the {\em SearchPath}
\STATE Create the {\em ConditionalPath}
\STATE Create the {\em ReturnPath}
\STATE Evaluate the query
\ENDIF
\IF {$\exists$ {\em SearchPath} in {\em PathDictionary}}
\STATE $Q_{B} \leftarrow$ {\em ConditionalValue}
\ELSE
\STATE Display {\em Not-Found-Element}
\ENDIF
\end{algorithmic}
}
\end{algorithm}

Once defined the concept query $Q$, it is inserted into the concept $T(C_{\leq})$  by using the method of incremental construction of Godin \cite{proceeding10}. The generalized view obtained corresponds to the concept lattice motion is noted $T_{\oplus}(C_{\leq}\oplus Q)$ where $C_{\leq}\oplus Q$ is the new set of concepts resulting from the insertion of the application in $T(C_{\leq})$.

We consider all nodes in the initial generalized view $T(C_{\leq})$.  The insertion of the query concept $Q$ follows the following properties.

\begin{property}
We consider all nodes in the initial generalized view $T$. We add the new element $Q$ to all extensions of nodes representing the search path.
\end{property}
\begin{property}
The new concepts are of form $(A\bigcup{Query},~B\bigcap f({Query}))$  for some concepts $(A,B)$. So, we have new sets $A^{'}$ and $B^{'}$. In this case, the concept is called a generator for the new pairs.
\end{property}
\begin{property}
  Each new item in the set $A^{'}$ in $V$ is the result of the intersection of $f(X)$ such as $x\subset X$ with a $B^{'}$ already present in $O$.
\end{property}
\begin{property}
\textbf{Property 4.} The son's concepts of former concepts doesn't change. The generator is also the only pair that becomes old son a new pair. A new pair may have a son but it is also a new concept.
\end{property}
\begin{property}
Parents of older concepts that do not generate new concepts remain unchanged. In addition, parents of modified concepts do not change.
\end{property}
The goal is to generate $T_{\oplus}(C_{\leq}\oplus Q)$ from $T(C_{\leq})$ and changing $X$ and $X^{'}$. The new concepts are derived from concepts of generators within properties mentioned above.
\subsection{Searching Query Answers}
Once the concept is inserted into the generalized view, the course of this structure can begin to search answers.
\begin{property}
  An object $o$ is relevant for a given query $Q=(Q_{A},Q_{B})$ if and only if it is characterized by at least one of the data $Q_{B}$.
\end{property}
Given a query $Q=(Q_{A},Q_{B})$, all relevant objects are within reach of $Q$ and their upper bound in generalized view $T_{\oplus}(C_{\leq=}\oplus Q)$ since the intent of each of these concepts is included in $Q_{B}$.

Let be consider $R_{o}(Q,C)$ all relevant objects for query $Q$  considered in the set of formal concepts $C$. Intuitively, the search algorithm of relevant objects, try to insert the query concept in $T(C_{\leq})$ to produce $T_{\oplus}$. Then, all objects appear in the extension of $Q$ in $T_{\oplus}$ are inserted into the list.

The following algorithm gives the all steps of this process.\\
\begin{algorithm}[h!]
\small{
\caption{Searching query answers}
\label{alg5}
\begin{algorithmic}[1]
\REQUIRE A query $Q=(Q_{A},Q_{B})$ where $Q_{A}=\emptyset$ and the generalized view $T(C_{\leq=})$
\ENSURE $T_{\oplus}(C_{\leq}\oplus Q)$ and a set of answers $R_{o}(T_{\oplus},C,level)$
\STATE Build the concept $Q=(Q_{A},Q_{B})$
\STATE Insert $Q$ in $T(C_{\leq})$
\STATE Search in $T_{\oplus}$ the new concept $Q=(Q_{A}\bigcup Q_{B}^{'})$
\STATE $level=0$
\STATE $UpperBounds(Q,C,level)\leftarrow {Q}$
\STATE $R_{o}(T_{\oplus},C,level)\leftarrow \emptyset$
\REPEAT
\FORALL {$C=(A,B)\in maj(Q,C,level)$}
\IF {$B\neq \emptyset$}
\STATE $R_{O}(T_{\oplus},C,level) \Leftarrow R_{O}(T_{\oplus},C,level) \cup A$
\ENDIF
\ENDFOR
\STATE $level \Leftarrow level+1$
\UNTIL $maj(Q,C,level)\Leftarrow \emptyset$
\end{algorithmic}
}
\end{algorithm}
\section{Conclusion}
The need for a mining model for XML documents becomes important. So, the principal idea of this work is to propose a FCA-based model for indexing and mining both XML structure and content.

For this reason, we have proposed a FCA-based model which aims is to ensure both the indexing and the querying of XML data while achieving a data conceptual classification.

The recourse to the FCA, as a solid mathematical foundation, is proved its efficiency in the indexing process of XML documents.

The aim of this process is to facilitate the querying one while ensuring the XML tree traversal, the conceptual classification and the conceptual scaling by generating a nested-based structure.

After indexing both structure and content, the querying process consists of generalizing concepts lattices into a generalized view. After that, a concept XQuery query is defined to be able to insert it in this view.

 Then, the coursing process began which permits the searching of answers following user's queries.

As future work, we propose to i) compare our FCA-based model with other classic mining models, ii) extend this model on a flexible querying process while using fuzzy predicates, iii) extend this model for querying multi-structured documents and finally iv) exploit this approach to facilitate querying  XML data and data warehouse, in which a native XML DB stores data and performs multi-dimensional OLAP (On-Line Analytical Processing queries).


\begin{thebibliography}{4}
\bibitem{book1}Bradely, N.: The XML Companion. Paperback, Subsequent Edition (2001).
\bibitem{url1}Extensible Markup Language (XML) Recommendation, \url{http://www.w3.org/TR/2000}.
\bibitem{book4}Tagarelli, A.: XML Data Mining: Models, Methods, and Applications. University of Calabria, Italy (2011).
\bibitem{proceeding1}Luk, R. W., Leong, H., Dillon, T.S., Shan, A.T., Croft, W.B., Allan, J.: A survey in indexing and searching XML documents. In: Journal of the American Society for Information Science and Technology, vol. 53(3), pp.415--435 (2002).
\bibitem{proceeding2}Weigel, F., Meuss, H., Bry, F., Schulz, K. U.: Content-aware data-guides: Interleaving IR and DB indexing techniques for efficient retrieval of textual XML data. In: European Conference on Information Retrieval, pp. 378--393. Sunderland, UK (2004)
\bibitem{proceeding3}Li, Q., Moon, B.: Indexing and querying XML data for regular path expressions. In: Proceedings of the $27^{th}$ International Conference on Very Large Databases Conference, Roma, Italy (2001)
\bibitem{book2}Ganter B., Wille R.: Formal concept analysis. Mathematical foundations. Springer Verlag, Berlin (1999).
\bibitem{proceeding4}Wille, R.: Restructuring lattice theory: An approach based on hierarchies of concepts. Ordered sets, vol. 23, Rival Editor, pp. 445--470 (1982).
\bibitem{book3}	Ganter, B., Wille, R.: Formal concept analysis, Mathématical fondations. Springer Verlag, Berlin (1999).
\bibitem{proceeding5} Ayadi, D., Arfaoui, O., Sassi-Hidri M.: Using conceptual scaling for indexing XML native databases. In: The $4^{th}$  International Workshop on XML Data Management, pp. 309--318 (2012).
\bibitem{proceeding6}Li, Q., Moon, B.: Indexing and Querying XML Data for Regular Path expressions. In: Proceedings of $27^{th}$ International Conference on Very Large Databases, pp. 361--370. Rome, Italy (2001).
\bibitem{proceeding7}Kyu, Y., Yoo, S., Yoon, K.: Index Structures for Structured Documents. In: IEEE International Conference on Development and Learning, pp. 91-99 (1996).
\bibitem{proceeding8}Valtchev, P.: An algorithm for minimal insertion in a type lattice. Computational Intelligence, vol. 15(1), pp. 63-78 (1999).
\bibitem{proceeding9}Chamberlin, D.: XQuery: An XML Query Language. IBM Systems Journal, vol. 41(24), pp. 597--615  (2002).
\bibitem{proceeding10}Godin, R., Missaoui, R., Alauoi, H.: Incremental Concept Formation Algorithms Based on Galois (Concept) Lattices. Computational Intelligence, vol. 11, pp. 246--267 (1995).
\end{thebibliography}
\end{document}